\documentclass[12pt]{iopart}
\begin{document}
 \begin{flushright}
 NIKHEF/99-032 \\
 \end{flushright}
\jl{4}

\title[Heavy Flavours at Colliders]{Heavy Flavours at Colliders}

\author{Eric Laenen\dag}

\address{\dag\ NIKHEF Theory Group, Kruislaan 409, 1009 DB,
Amsterdam, The Netherlands}

\begin{abstract}
I review some topics in the production and decays 
of heavy flavours that are relevant for collider physics. 
In particular, I discuss the present status and 
some recent progress related to masses, 
parton densities and fragmentation functions of heavy quarks, 
as well as threshold resummation, polarized onium production at high
transverse momentum, and a factorization theorem for $B \rightarrow\pi\pi$ decays.
\end{abstract}



\section{Introduction}

When we consider the main physics goals of present and upcoming 
high energy colliders, we may justly distinguish the heavy quarks 
charm, bottom and top from their light brethren up, down and strange
as being especially important.

In the Standard Model (SM), on the electroweak side, the ability to 
individually identify (tag) various types of heavy-flavoured hadrons
allows detailed scrutiny of its family structure,
via e.g. measurements of the CKM matrix elements, couplings
to the $Z$-boson, production asymmetries etc.
The large Higgs Yukawa couplings of the heavy quarks
are central to many Higgs search strategies.
On the QCD side, this same ability gives us a closer look into 
hard scattering production mechanisms than more average quantities 
like jets. Moreover, the fact that
their masses are larger than the scale where the strong
coupling regime in QCD begins means that physics
involving these heavy quarks is much more amenable to perturbative
analysis. 

Concerning physics beyond the SM,
experience with SM heavy flavour production helps in
predicting production rates for non-SM heavy particles.
Also, certain B-decays (e.g. $B\rightarrow X_s\gamma$) are 
sensitive to the virtual presence of non-SM particles
(see e.g. \cite{Buras:1998ra} and references therein).

Thus there is a large arena in which heavy flavours play a key role. 
I will discuss a QCD-biased selection of topics
that are chosen for being fundamental to (heavy flavour) physics, 
and for either recently having undergone significant progress,
or for showing clear potential for such. Consequently,
what follows is somewhat of an itemized list, rather than an story with a plot.

\section{Heavy Quark Mass}

Central to all heavy flavour physics at high energy
colliders is the need for knowing their masses
accurately. But the concept of mass for a
confined, unstable fermion needs careful definition
and consideration. Let us review 
some definitions and discuss their relations.

\subsection{Pole Mass $M$}

\def\slash#1{\rlap{\hbox{$\mskip 1 mu /$}}#1}      
\def\Slash#1{\rlap{\hbox{$\mskip 3 mu /$}}#1}      
\def\SLash#1{\rlap{\hbox{$\mskip 4.5 mu /$}}#1}    
\def\SLLash#1{\rlap{\hbox{$\mskip 6 mu /$}}#1}      

This is the most common and natural definition of the electron or muon mass 
in QED.
In QCD, the heavy quark pole mass is defined via the inverse heavy quark Feynman propagator
\begin{equation}
  \label{eq:1}
 S_F^{-1}(p) =  \slash{p}-m_0-\slash\Sigma(p,g_0,m_0)\,,
\end{equation}
where $\slash{\Sigma}$ is the sum of all one-particle
irreducible QCD corrections to the propagator,
$m_0$  is the bare heavy quark mass, and $g_0$ is
the bare coupling. For this discussion
I ignore electroweak corrections which give the 
heavy quark a finite width.
The pole mass $M$ is defined by the requirement that
the propagator has a pole at $p^2 = M^2$. It is related
to the bare mass by 
\begin{equation}
  \label{eq:6}
  M = Z_m m_0\,,
\end{equation}
with the UV-regulator dependent $Z_m$ a series
in $g_0$. To determine this series,
one puts the right hand side of 
(\ref{eq:1}) to zero at $p^2 = M^2$, and solves for $M$ iteratively.
UV divergences in $Z_m$ are cancelled by those of the
bare mass and coupling.
Thus defined, $M$ is also infrared-finite and gauge-independent
to all orders \cite{Tarrach:1981up,Kronfeld:1998di}.

There are a number of attractive features to this definition,
besides its correspondence with QED and its intuitive nature:
it is essentially the pole mass that is directly 
reconstructed from the heavy quark
decay products, it can be calculated not just in full
QCD, but also in effective theories such as 
non-relativistic QCD (NRQCD) and heavy-quark effective
theory (HQET), and it is the required definition
in renormalization schemes \cite{Collins:1978wz} in
which heavy particles manifestly decouple in Green's
functions with small external momenta.

There is also a problem with the pole mass definition.
A renormalon analysis \cite{Beneke:1994sw,Bigi:1994em}
shows that the pole mass of a heavy quark 
has an intrinsic uncertainty of order $\Lambda_{\rm QCD}$.
A heuristic argument for this intrinsic uncertainty, valid also if the 
heavy quark has a finite width, may be given based on confinement \cite{Smith:1997xz}:
if the pole mass $M$ could be known to arbitrary precision,
then the closer to $M^2$ the (momentum)$^2$ of a produced heavy quark would be, 
the longer-lived it would be, ultimately
leading to it being freely observable. But because we
know quarks are confined, it must be impossible
to know the pole mass to a precision greater
than the confinement scale $\Lambda_{\rm QCD}$.

For the top quark this ambiguity has no impact
on any physics analysis, since the top decays long
before any confinement effects come into play
\cite{Bigi:1986jk,Fadin:1987wz,Fadin:1988fn}.

This uncertainty does however matter for the
charm and bottom masses, and for precise 
mass determinations other definitions are required.
There is a particular need for a precisely
known heavy quark mass in heavy meson decay rates,
which scale as $({\rm mass})^5$ and
are needed to infer the CKM element $V_{cb}$.

\subsection{$\overline{\rm MS}$ mass $m(\mu)$}

One may also treat the bare mass $m_0$ in the
QCD Lagrangian as a two-particle coupling constant, which 
upon UV $\overline{\rm MS}$ renormalization, 
\begin{equation}
  \label{eq:7}
  m(\mu)=z_m(\mu)m_0
\end{equation}
becomes a running $\overline{\rm MS}$ mass $m(\mu)$.
This more formal mass definition lacks the intuitive
character of the pole mass, but does not suffer
from non-perturbative ambiguities. It is thus
a better definition for precision measurement.

Both mass definitions can be viewed as different
renormalizations of the same bare mass, so one 
may express $m(m)$ in $M$. This relation 
reads to N$^3$LO \cite{Gray:1990yh,Chetyrkin:1999qi} 
\begin{eqnarray}
  \nonumber
\frac{M}{m(m)}&=&1+1.333 \Bigg(\frac{\alpha_s(m)}{\pi}\Bigg)
 +(13.44-1.041 n_l) \Bigg(\frac{\alpha_s(m)}{\pi}\Bigg)^2
\\\nonumber&+&
(194(5)-27.0(7)n_l+0.653 n_l^2) \Bigg(\frac{\alpha_s(m)}{\pi}\Bigg)^3
+ {\cal O}(\Lambda_{\rm QCD})
\end{eqnarray}
with $n_l$ the number of light flavours, and the numbers in 
brackets in the last term reflect an uncertainty
in Pad\'{e} approximations used to obtain
the $ {\cal O}(\alpha_s^3)$ \cite{Chetyrkin:1999qi} results.
Very recently, the relation also been computed exactly 
\cite{Melnikov:2000qh}.

\subsection{Other mass definitions}

There has been much recent development in determining
the heavy quark masses from $Q\bar{Q}$ systems near threshold.
Their properties may be computed in an effective 
theory, non-relativistic QCD (NRQCD) \cite{Bodwin:1995jh}.
Observables in these systems are in general less infrared-sensitive
than the pole mass, so that this mass definition is not preferred.

To determine $m(m)$ accurately, other,
intermediate mass definitions or schemes are very valuable.
In \cite{Beneke:1998rk} the Potential Subtraction (PS) scheme was
proposed. It was shown in \cite{Beneke:1998rk,Hoang:1998nz} 
that the nonrelativistic Coulomb $Q\bar{Q}$ potential 
$V(r)$ has precisely the same infrared sensitivity as $-2M$.
The combination $2M + V(r)$ occurs naturally in 
the process $e^+e^-\rightarrow Q\bar{Q}X$ near-threshold,
which is governed by the Schr\"{o}dinger operator $H-E$, with
\begin{equation}
  \label{eq:9}
H =  \frac{-\nabla^2}{2M} + V(r),\qquad E = \sqrt{s}-2M\,.
\end{equation}
This suggests to redefine a potential and mass both
with much less infrared sensitivity
\begin{equation}
  \label{eq:11}
V(r,\mu_f) = V(r)+2\delta m(\mu_f),\;\;
m_{PS}(\mu_f) = M-\delta m(\mu_f) \,,
\end{equation}
with $\mu_f$ a subtraction scale,
and $2\delta m(\mu_f)$ the IR-sensitive part of the
potential. Solutions to 
the Schr\"{o}dinger equation may be expressed in
$m_{PS}(\mu_f)$, but are as a whole independent of $\mu_f$.
Comparing with data leads to a determation of
$m_{PS}(\mu_f)$, at $\mu_f = 2$ GeV e.g.
The relation of $m_{PS}(\mu_f)$ to $m(m)$ is 
known to order $\alpha_s^3$ \cite{Chetyrkin:1999qi,Schroder:1998vy}.

Other schemes, e.g. based on the $\Upsilon(1S)$ mass
\cite{Hoang:1998ng}, and on the B-meson mass
\cite{Bigi:1997fj} exist.
An extensive review of
these issues together with a detailed comparison between various
schemes, as well as results for the heavy quark masses 
can be found in \cite{Beneke:1999fe,Beneke:1999zr}.

The $\overline{\rm MS}$ masses of the heavy quarks are
at present known to about 4\% for the case of top
\cite{Abe:1995hr,Abachi:1995iq} (from direct reconstruction), 
to within 2\% for the case of the bottom \cite{Beneke:1999fe}
(from sum rules), i.e. the error is already much
less than $\Lambda_{\rm QCD}$, and
to within 10\% for the case of charm \cite{Pineda:1998ja}
(extracted from the charmonium spectrum).

\section{Variable Flavour Number Schemes}

It is a subtle issue to choose a model for heavy flavour 
production in collisions with initial hadrons,
such as deep-inelastic scattering.
One must decide whether one wants to employ a 
fixed number of light flavours in the evolution
of the parton densities and $\alpha_s$ and treat the
heavy flavour as an external quantum field: the ``fixed flavour-number
scheme'' (FFNS); or whether one wishes to change the 
number of dynamical flavours depending on the scale present in the
parton subprocess, and introduce heavy quark parton densities: 
the ``variable flavour-number scheme'' (VFNS).
The past few years have seen much progress in the
development of such schemes, which began in \cite{Aivazis:1994pi}.
Their development is leading to a considerable
increase in understanding field theoretic treatments
of heavy quarks in high energy scattering.

Although the FFNS is more straightforward in use, and
has been the scheme of choice for almost all available
NLO calculations of heavy flavour production, it can
suffer from large perturbative corrections 
proportional to $\alpha_s\ln(Q^2/m^2)$ when the 
hard scale $Q$ of the reaction is significantly larger than
the heavy quark mass $m$. In a VFNS one views such terms 
as collinear divergences, which one may consistently factorize into 
the set of parton densities at hand, augmented by one for the heavy quark.
An all order proof for this was given in \cite{Collins:1998rz}.
In so doing one may control such logarithms to all orders, without 
relinquishing control over the region $Q\simeq m$. 

From an effective field theory viewpoint, one must match
a theory with $n_f=n_{l}+1$ dynamic quark species,
to one with only $n_{l}$ dynamic ones,  
in which the heavy quark is not treated as a parton. 
The coefficient functions of the former are 
the residues of the latter after subtracting all
$\ln(\mu^2/m^2)$ terms via the appropriate
operator matrix elements \cite{Buza:1996ie}.
Their absorption into the $n_l$ set of parton densities
leads to matching conditions between these two sets of densities,
(as well as the QCD coupling $\alpha_s(\mu)$)
at a matching scale, which is conveniently (but not
necessarily) chosen to be the heavy quark mass $m$.
These conditions are now known for $\alpha_s$, in the $\overline{\rm MS}$ scheme,
to N$^3$LO \cite{Bernreuther:1982sg,Chetyrkin:1998un,Larin:1995va}
and for the $\overline{\rm MS}$ parton densities to NNLO \cite{Buza:1996ie}.
Both quantities develop a discontinuity at the matching point from NNLO onwards.

Various VFNS implementations now exist 
\cite{Aivazis:1994pi,Buza:1998wv,Thorne:1998ga},
mostly in the context of deep-inelastic heavy
quark production, i.e. for the structure functions 
$F_{2,L}^{\rm charm}$.
They differ mainly in the order in 
$\alpha_s$, in the treatment of the matching
conditions, and in the extent to which finite mass effects are retained
in the coefficient functions. The latter are relevant for
the formation of actual thresholds in the partonic
cross sections. 

In \cite{Chuvakin:1999nx} a NNLO VFNS scheme
was constructed for $F_{2,L}^{\rm charm}$
with ${\cal O}(\alpha_s^2)$ matching conditions and
full mass dependence in all coefficient functions. 
It was compared with a NNLO VFNS scheme in which the subtraction
terms, combinations of matrix elements and lower order
coefficient functions, have only logarithmic mass dependence \cite{Buza:1998wv}.
Incidentally, not that in a VFNS in which logarithms of $m$ are viewed as 
divergences so $F_{2,L}^{\rm charm}$ are no longer infrared-safe observables, 
since mixing occurs with the $F_{2,L}^{\rm u,d,s,g}$ from ${\cal O}(\alpha_s^2)$
onwards. Jet definitions or fragmentation functions are required
(a $Q\bar{Q}$ invariant-mass cut was used in \cite{Chuvakin:1999nx}).
The two schemes give essentially the
same result for $F_2^{\rm charm}$.
Both tend to the three-flavour FFNS results
at low scale, and to a four-flavour massless-charm result at high
scale. The NNLO charm density was found to start off negative at small
$x$ at the matching scale (a consequence of the
NNLO matching conditions). In \cite{Chuvakin:1999ip} 
four- and five-flavour NLO and NNLO PDF sets were constructed from the
GRV98 \cite{Gluck:1998xa} three-flavour set by using the
NNLO matching conditions. 

The practical necessity of using VFNS schemes 
in understanding present collider data is not yet clear
(e.g. deep-inelastic charm production at HERA is
well described by a three-flavour number scheme
for all scales). But if we wish to have accurate
predictions for heavy flavour production cross sections
for future high energy colliders,
free from large logarithms in their perturbative
series, further development of higher-order VFNS 
implementations is very important.

\section{Threshold Resummation}

Semi-inclusive hadronic cross sections factorize in perturbative QCD 
into universal, non-perturbative parton distribution functions
and fragmentation functions, summarizing long-distance dynamics, and 
perturbatively calculable hard scattering coefficient functions,
which organize all short-distance effects. 
However, sizable long distance effects remain at higher orders
in the hard scattering functions even after infrared divergences have cancelled.
These so-called Sudakov corrections assume the form of double-logarithmically enhanced
distributions that become singular at partonic threshold.
The goal of threshold resummation is to resum these corrections
to all orders in perturbation theory, thereby regaining control over
the perturbation expansion, a program that has been advanced considerably
in the last couple of years. See \cite{Kidonakis:1999ze} for a recent review.

Striving for accuracy in this context means resumming 
not only leading logarithmic (LL) corrections, but
also next-to-leading ones and so on. How to achieve this
for processes that are electroweak at lowest order, 
such as Drell-Yan, has been known for a while
\cite{Sterman:1987aj,Catani:1989ne}. 

It has recently been understood how to 
do this for general QCD processes, of which
the Born amplitudes may comprise various colour structures. 
For example, in the production of a heavy quark-antiquark
pair in a gluon-gluon collision, the final state 
has the SU(3) representations
${\bf 3}$ and ${\bf \bar{3}}$ 
and the initial state two octets. Denoting the
heavy quark colour labels by $k$ and $\bar{l}$ 
($k,\bar{l}=1,\ldots 3$) and
the gluon colour labels by $a,b$ ($a,b=1,\ldots,8$)
the external particles in the Born amplitude can
be colour-coupled in three ways 
\begin{equation}
  \label{eq:cscoup}
\delta_{ab} \delta_{k\bar{l}},\qquad
f_{abc} (t^c)_{k\bar{l}},\qquad
d_{abc} (t^c)_{k\bar{l}}\,.
\end{equation}
The Sudakov corrections are expressed in terms of 
functions that are singular at partonic threshold,
but give (large) finite results when integrated
against smooth functions such as parton densities.
They occur in general for each of these colour structures.
To resum them to NLL accuracy and beyond
requires understanding how these structures mix under soft gluon radiation. 
\cite{Kidonakis:1996aq,Kidonakis:1998bk,Bonciani:1998vc}. 
Correspondingly, the all-order threshold-resummed cross section 
for this channel is a 3 by 3 matrix. 

In general, threshold resummation is based on a 
refactorization of a partonic cross section near threshold
\cite{Sterman:1987aj,Contopanagos:1997nh,Collins:1989bt}
into various classes of functions, or effective
theories: the double logs are resummed in a colour-diagonal theory 
for high-momentum collinear incoming partons moving in particular direction, and a similar
one for outgoing ones; the single-logarithmic colour-coherence 
effects are resummed in a colour-mixing theory for soft (eikonal) gluons.
The matching to full QCD happens via hard, far off-shell partons. 
Evolution equations in these effective theories 
are given in \cite{Sterman:1987aj,Contopanagos:1997nh,Kidonakis:1998nf}.
How to contruct in general NLL resummed cross sections in phenomenologically
common one-particle inclusive kinematics was explained in \cite{Laenen:1998qw}. 

Various threshold-resummed heavy quark production cross sections have thus
been built and phenomenological studies performed. 
Earlier partially-NLL threshold resummations results
for top quark production at the Tevatron were 
performed \cite{Laenen:1994xr,Berger:1996ad,Catani:1996yz}.
For heavy quarks, full NLL resummations now exist 
for hadroproduction \cite{Kidonakis:1996aq,Bonciani:1998vc,klmv}
and electroproduction \cite{Laenen:1998kp}.

Besides having intrinsic value for providing more insight in
the nature of perturbative cross sections in near-elastic
regions of phase space, there are quite 
practical benefits to threshold resummation: the ability
to assess or predict, in appropriate
kinematic regions, the higher order corrections to the
cross section \cite{Laenen:1998kp,klmv}, 
and a general reduction in the scale dependence of resummed cross sections
\cite{Sterman:1999gz,Bonciani:1998vc,Catani:1998tm,Catani:1999hs}.

\section{Heavy Quark Fragmentation}

Fragmentation functions (FF's) match parton production cross sections
to ones in which a specified hadron, such as a heavy-flavoured
meson or baryon, is produced. As such they are in principle as essential
as parton densities in the comparison of QCD calculations to data.
Heavy quark FF's are special for two reasons. First, since QCD cannot change flavours,
there is a direct link between the flavour of a produced heavy meson 
and a produced heavy quark, so that we may flavour-tag heavy quarks.
Second, collinear divergences associated with
radiation from the hard parton are screened by the heavy quark
mass, hence, loosely speaking, a large part of the fragmentation
process is actually perturbative. In the last few years
(perturbative) heavy quark fragmentation has reached a quite mature understanding,
and has now been applied in its fully developed form 
to a few heavy quark production cross sections. 

For the description of the nonperturbative transition of a heavy quark 
($Q$) to a heavy hadron ($Q\bar{q}$), the 
Peterson fragmentation function \cite{Peterson:1983ak} is often still
adequate. It contains in essence 
one parameter, the mass ratio squared $\epsilon = m_{\bar{q}}^2/m_Q^2$.
But heavy quark FF's have a perturbative component as well. 
The perturbative FF (PFF) satisfies the DGLAP \cite{Altarelli:1977zs,Gribov:1972ri,Dokshitzer:1977sg}
evolution equation:
\begin{equation}
  \label{eq:8}
\frac{d D_{i,{\rm pert}}(x,\mu)}{d\ln\mu}
 = \sum_j \int_x^1 \frac{dz}{z} P_{ij}(\frac{x}{z},\alpha_s(\mu))
D_{j,{\rm pert}}(z,\mu) \,,
\end{equation}
where $i,j$ label parton flavours. With an initial condition at $\mu_0\simeq m$, 
(\ref{eq:8}) determines the PFF at $x,\mu$, and resums $\ln(\mu/\mu_0)$ to
all orders. This initial condition for the PFF  was first computed in
\cite{Mele:1991cw} and is, for $i=Q$
\begin{eqnarray}
  \label{eq:10}
\fl D_{Q,{\rm pert}}(z,\mu_0) = \delta(1-z)+\frac{\alpha_s(\mu_0)C_F}{2\pi} 
\left[\Bigg(\frac{1+z^2}{1-z}\Bigg)\Bigg(\ln(\frac{\mu_0^2}{m^2})-1-2\ln(1-z)\Bigg)\right]_+\,.
\end{eqnarray}
For heavy quark production at $p_T \gg m$, the choice $\mu=p_T$ 
in the solution of (\ref{eq:8})
then resums the $\ln(p_T/m)$ logarithms to all
orders. The full fragmentation function requires combining
the perturbative FF with a non-perturbative part modelling
the final hadronization.

Heavy flavour production in  $e^+e^-$ collisions
has been thoroughly studied both at fixed order 
\cite{Nason:1997nw,Rodrigo:1999qg,Bernreuther:1997jn}
and in a combined fixed order plus next-to-leading logarithmic 
(here $\ln(E/m)$ with $E$ the beam energy) resummed approach
\cite{Nason:1999zj}, using the Peterson function
for the non-perturbative transition. This combined
approach is actually a VFNS scheme for heavy quark fragmentation.
Increasing the order in perturbation theory turns out \cite{Nason:1999zj} 
to reduce the Peterson parameter $\epsilon$, 
which corresponds to a harder non-perturbative fragmentation function.
The importance of ${\cal O}(m/E)$ terms in the fixed-order
part of the calculation is found to be minor.

This formalism has been applied as well
to the Tevatron $b$-quark $p_T$ cross section,
for which the data exceed the central NLO-theory estimate by a factor of two,
in \cite{Cacciari:1994mq} and more recently
in \cite{Cacciari:1998it}. 
At large $p_T$ the theoretical uncertainty due to scale
variations was found to be reduced with respect
to the fixed-order approach, 
but the cross section decreased. At moderate $p_T$, the cross section 
gets somewhat enhanced, but not enough to explain the data-theory discrepancy.
Another study involving FF's in heavy quark hadroproduction,
in the context of the ACOT VFNS, was performed in \cite{Olness:1997yc}.
A full VFNS implementation with all available knowledge about
heavy quark parton densities and fragmentation functions
included would be interesting. 
I note that the PFF formalism was also applied to $\gamma p$ \cite{Cacciari:1996fs} and
$\gamma \gamma$ \cite{Cacciari:1996ej} charm production. Similar conclusions 
were reached as for $p\bar{p}$ $b$-quark production. 

A puzzling situation has arisen in
charm electroproduction and photoproduction at HERA.
The data are well-described by NLO QCD plus
Peterson fragmentation, except at
low $p_T$ and large rapidity
It is surmised that remnant-beam
drag effects, which have been 
modelled in the string fragmentation model \cite{Norrbin:1998bw}, 
may exert influence here.
It is important to understand this effect,
as it affects $B-\bar{B}$ asymmetries at e.g. HERA-B
\cite{Norrbin:1999by}, and thus CP violation measurements.
An open problem is how to match the string
model for beam drag to leading twist QCD.

\section{Polarized Onium Production at the Tevatron}

Non-relativistic QCD (NRQCD) \cite{Bodwin:1995jh} 
is the effective field theory for light quarks, gluons and 
heavy quark-antiquark bound states (onia).
This effective theory allows besides an expansion
in $\alpha_s$, a systematic expansion in $v$, 
the heavy quark velocity inside the onium bound state.
Besides the QCD confinement scale $\Lambda$, 
and the heavy flavour mass $m$, 
the inverse onium size $mv$ and binding energy $mv^2$ are important scales
in this theory.

A few years ago a Tevatron measurement \cite{Abe:1992ww}
of the production rates of the $J/\psi,\psi'$ states revealed
an enormous excess over the prediction
of the colour-singlet model \cite{Schuler:1994hy}.
The measurement may however be understood in a NRQCD framework
\cite{Braaten:1995vv}.
Onia production at the Tevatron is described in NRQCD by
\begin{eqnarray}
  \label{eq:4}
\fl d\sigma(p+\bar{p}\rightarrow \psi(P,\lambda)+X) =
\sum_{ij} f_i\otimes f_j \otimes \sum_n
d\sigma(i+j\rightarrow Q\bar{Q}[n]+X) \langle O^{\psi(\lambda)}_n \rangle\,,
\end{eqnarray}
where the sum is over all allowed states, labeled by the spectroscopic notation
$^{2S+1}L_J$. Note the colour label $({\bf 1},{\bf 8},\ldots)$.
E.g. $n = ^3S_1^{({\bf 8})}, ^2S_0^{({\bf 8})}, ^3P_J^{({\bf 8})}$.
The operators $O^{\psi(\lambda)}_n$ are ordered in the expansion
parameters $\alpha_s$ and $v$. 
The content of this expression can be stated in
physical terms as follows. The onium production time scale 
is of order $1/m$, whereas
the much longer onium binding time scale is of order $1/(m v^2)$.
The physics of the binding is encoded
in the operator matrix elements.
The fact that coloured channels, in particular the
octet ones, are thus open to scatter into,
leads to large increase in the cross section
\cite{Braaten:1995vv}, allowing an explanation of
the Tevatron data. 

A test of the importance of these 
colour-octet states is the polarization of
the vector onium octet-state produced, $J/\psi$ and $\psi'$
\cite{Cho:1995ih, Cho:1995gb,Beneke:1996yb}.
A heuristic argument says that at large $p_T/m$,
this state results in essence from the fragmentation
of a nearly on-shell and transverse gluon,
leading to a transversely polarized onium state,
at large $p_T/m$, as soft-gluon binding effects
do not flip the spin. The polarization $\alpha$
of the produced onium may be inferred from
the angular distribution of the lepton pair
into which it decays:
\begin{equation}
  \label{eq:2}
\frac{d\Gamma}{d\cos\theta} \propto 1 + \alpha\cos^2\theta\,,
\end{equation}
where $\theta$ is the angle between the $\mu^+$ (say) momentum in the 
onium rest frame and the onium momentum in the lab frame, and
with $\alpha=-1,0,1$ corresponding to a
longitudinally polarized, unpolarized and 
transversely polarized onium state, respectively.
A careful theoretical assessment of the
expected polarization, based
on (\ref{eq:4}), was made in \cite{Beneke:1997yw}.

A very recent CDF measurement \cite{Cropp:1999ub}
of the $\psi(2S)$ polarization versus its transverse momentum
indicates no preference for $\alpha \rightarrow 1$ as $p_T$ increases!
But before we get really worried we should wait for more Tevatron data.
Additionally, lepto- and photoproduction data from HERA may shed
more light \cite{Beneke:1998re,Fleming:1997fq}.

\section{Factorization for $B\rightarrow \pi\pi$ decays}

Establishing CP violation in B-meson decays is 
one of the main physics goals at present and future
colliders. In the SM, CP violation
is accommodated via a complex phase in the CKM matrix, which is 
possible by having 3 families.
The intention is not just to see that CP is violated in B-decays,
but also to determine all entries of the CKM mixing matrix, expressible
in four angles $\alpha,\beta,\chi,\chi'$, with $\gamma\equiv\pi-\alpha-\beta$.
See e.g. \cite{Kayser:1996sv} for their definitions.

One of the main methods thereto is the
measurement of the time-dependent asymmetry 
\begin{equation}
  \label{eq:3}
  {\rm Asym}(t) = \frac{\Gamma_{B_d \rightarrow {\rm f}}(t) - 
\Gamma_{\bar{B}_d \rightarrow {\rm f}}(t)}{
\;\;"\;\; + \;\;"\;\;}
\end{equation}
A very important set of decay channels 
$\rm f$ are the two-body hadronic states 
$J/\psi K_s$, $\rho^+ \pi^-$, $\pi^+ \pi^-$, denoted collectively
$\pi \pi'$ for short.
In the simplest case, if the decay is dominated by one diagram and the final
state is a CP eigenstate, then
\begin{equation}
  \label{eq:two}
{\rm Asym}(t) \propto \sin(2\alpha)
\end{equation}
But when there is more than one diagram contributing,
leading to more interference terms
in $|A(B_d \rightarrow {\rm f})|^2$, 
the disentanglement of CP-violating weak phases
from strong phases is complicated. Hence calculational 
control over hadronization and strong phases is very desirable.
An important new development is a factorization of 
$B\rightarrow \pi\pi$ amplitudes \cite{Beneke:1999br},
allowing their calculation from first principles.
(QCD factorization for $B\rightarrow D\pi$ decays
was discussed in \cite{Dugan:1991de}).

$B \rightarrow \pi\pi'$ decays are described by 
the effective Hamiltonian \cite{Buchalla:1996vs}
\begin{equation}
  \label{eq:heff}
H_{\rm eff} = G_F \sum_{i} C_i\, O_i\,,  
\end{equation}
where $G_F$ is the Fermi constant, the $O_i$ are
$\Delta B=1,\,\Delta S=0$ four quark operators, and $C_i$ 
their Wilson coefficient functions. To compute the decay
rates and their asymmetries one must know the matrix elements
$\langle \pi\pi | O_i | B\rangle$. The proposed QCD factorization 
\cite{Beneke:1999br} for these amplitudes reads 
\begin{eqnarray}
  \nonumber
\langle \pi'\pi | O_i | B\rangle &=& f^{B\rightarrow\pi}
  \int dx T_i^I(x) \Phi_\pi'(x) \\  \nonumber
&+& \int d\xi dx dy\, T_i^{II}(\xi,x,y)\, \Phi_B(\xi)\Phi_\pi'(x)\Phi_\pi(y) \\  \nonumber
&+& {\cal O}
\Bigg(  \frac{\Lambda_{\rm QCD}}{m_b} \Bigg)
\end{eqnarray}
with $T_i^{I,II}$ calculable kernels, and the
$\Phi_{\pi,B}$'s light-cone distribution amplitudes.
The first term corresponds to the $B\rightarrow \pi$
form factor (see e.g. \cite{Weinzierl:1999sv,Khodjamirian:1997ub,Bagan:1997bp}
for its determination)
evaluated at the mass squared of the other pion ($\pi'$).
It benefits from the fact that soft gluons do not couple to a
small-size $q\bar{q}$ ($\pi'$) pair escaping from
the system. The second term is a ``standard'' Brodsky-Lepage 
\cite{Lepage:1980fj} term for exclusive scattering.
The analysis \cite{Beneke:1999br} shows that the
strong interaction phases for $B\rightarrow \pi\pi$
are ${\cal O}(\alpha_s)$ or 
${\cal O}( \Lambda_{\rm QCD}/m_b)$. If this theorem can be proven 
to all orders, the strong phases are calculable so that
CKM phases can be better extracted.

\section{Conclusions and Outlook}

We have just had a brief look at a few,
mostly QCD-related, topics in the 
physics of heavy flavours at colliders.
A common thread in the theoretical approach to these topics
is perhaps that of effective field theories.
The value of using such a framework is
in some cases very high. Thus, our grasp of $Q\bar{Q}$ systems has been revolutionized by
NRQCD. Likewise the physics of heavy-light system has
been understood much deeper in HQET, where new symmetries
are present. In other cases, taking an effective theory 
viewpoint is at present interesting, but not yet quite as revealing.

I hope to have left an impression of the substantial progress made 
in a variety of areas in heavy flavour physics, 
and that conceptual understanding in many cases is 
quite mature, but also to have conveyed a notion that much
is still left to explore.
With heavy flavours such an important component
to many analyses at future colliders, 
we may be confident that such explorations will take place.

I would like to end by thanking the workshop organizers and
conveners for providing for such a stimulating and truly 
enjoyable atmosphere, and the heavy flavour working group members for the 
pleasant and educational discussions.

\section*{References}


\end{document}